\documentstyle[12pt,epsfig]{article}
 \newcommand{\p}{{\bf p}}
 \newcommand{\bk}{{\bf k}}
 \newcommand{\bK}{{\bf K}}
 \newcommand{\0}{{\bf 0}}
 \newcommand{\lam}{{\lambda}}
 \newcommand{\ra}{{\rangle}}
 \newcommand{\la}{{\langle}}
 \newcommand{\be}{\begin{equation}}
 \newcommand{\ben}{\begin{eqnarray}}
 \newcommand{\een}{\end{eqnarray}}
 \newcommand{\ee}{\end{equation}}
 \newcommand{\po}{\stackrel{o}{\p}}
 \newcommand{\dsty}{\displaystyle}
 \newcommand{\Mo}{\stackrel{o}{M}}
 \newcommand{\pmmo}{\stackrel{\not\;\not{o}}{\p\smash{\lefteqn{_1}}}\phantom{1}}
 \newcommand{\pnno}{\stackrel{\not\;\not{o}}{\p\smash{\lefteqn{_2}}}\phantom{1}}
 
 \newcommand{\poc}{\stackrel{o}{ p}}
 \newcommand{\ponotc}{\stackrel{\not\,{o}}{p}}

 \newcommand{\pmo}{\stackrel{\not\,{o}}{\p\smash{\lefteqn{_1}}}\phantom{1}}
 \newcommand{\pno}{\stackrel{\not\,{o}}{\p\smash{\lefteqn{_2}}}\phantom{1}}

\newcommand{\Dp}{\stackrel{+}{D\smash{\lefteqn{^{(1/2)}}}}\phantom{aa1}}

\begin{document}

\begin{center}
{\bf
 CHARGE RADII OF THE PROTON AND VALENCE QUARKS
IN THE QUASIPOTENTIAL MODEL
}
\\
 T.P.Ilichova, S.G.Shulga
\\[6pt] {\it Francisk
Skaryna Gomel State University, LPP JINR, Dubna} \end{center}

\begin{abstract}
The mean square radius of the proton charge distribution
was studied in the framework of the relativistic
quasipotential quark model in assumption of the SU(6)-symmetry.
It was shown that the proton charge radius is represented as function
$f(m/\gamma)/M^2$ for point quarks
($m$ is a quark mass,
$\gamma$ is the scale of the coupling energy,
$M$ is the nucleon mass, $f$ is undimentional function, which
does not depended on $M$).
It was shown, that in the ultra relativistic region
the mean square radius of the bound system can have
negative value. To describe
simultaneously nucleon magnetic moments and proton
radius in the oscillator model
it is need to suppose that quark has the negative
mean square radius $\la r^2_{1q}\ra = -1.875/m^2 $.
\end{abstract}

\begin{center}
{\bf Introduction.}
\end{center}

\par
In works~\cite{IzvVuzov1_Ilicheva,IzvVuzov2_Ilicheva,IzvVuzov3_Ilicheva}
the formalizm of the quark
model in the quasipotential approach was considered:
the nucleon quark wave function was constructed
in the impulse approximation~\cite{IzvVuzov1_Ilicheva},
the calculations of the magnetic
moments~\cite{IzvVuzov2_Ilicheva} and the
ratio of the axial and vector coupling
constants~\cite{IzvVuzov3_Ilicheva} were
performed.
It was obtained that model has property
which is followed by the relativistic kinematics of quarks:
nucleon magnetic moments
(in units of nuclear magnetons)
and  ratio $G_A/G_V$ depend on one undimensional
parameter, which is constructed as ratio of the quark mass and
the energy scale parameter,
and don't depend on the ratio of the quark mass and nucleon mass.
In this case the valent qaurk mass and energy of coupling
are intrinsic parameter, which are related.
Nucleon charge radius, considered in the this paper, has
this property.

\begin{center}
{\bf 1. Basic relations of the model}
\end{center}
\par
In the quasipotential quark model~\cite{IzvVuzov1_Ilicheva}
the electromagnetic current in the rest frame of the
initial nucleon has the form~\cite{IzvVuzov2_Ilicheva}:
$$
\la \bK \lam'_J\lam'_T
\mid J_{\mu}(0)\mid\0\lam_J\lam_T
\ra \equiv
 e J_{\mu}^{\lam'_J\lam_J}(\bK,{\0})
 \delta_{\lam'_T\lam_T} =
$$
\be
= 3 e \int d\Omega_{\p_1} d\Omega_{\p_2}
     \varphi(\Mo'_0) G_{\mu} \varphi(M_0) \delta_{\tau'\tau},
\label{current}
\ee
where
\ben
 && \Mo'_0 \equiv E_{\po_1} + E_{\po_2} + E_{\po'_3},
 \qquad
    M_0 \equiv  E_{\p_1} + E_{\p_2} + E_{\p_3},
\nonumber\\
 && G_{\mu}^{(3)} \equiv
\frac{1}{ E_{\stackrel{o}{\p}'_3} E_{\p_3} }
(\chi^{\lam'_J\lam'_T} B^{1} B^{2} j^{(3)}_{\mu}
B^{3} \chi^{\lam_J\lam_T}),
\label{G_mu}\\
 && B^{(1)}_{\{\lam'_{s_k}\lam'_{t_k}\lam_{s_k}\lam_{t_k}\}} \equiv
\Bigl[
\Dp_{\lam'_{s_1}\lam_{s_1}}(L_{\po_{12}},\pmo)
\Dp_{\lam'_{s_2}\lam_{s_2}}(L_{\po_{12}},\pno)
\Bigr]
\delta_{\lam'_{t_1}\lam_{t_1}}
\delta_{\lam'_{t_2}\lam_{t_2}}
\delta_{\lam'_{t_3}\lam_{t_3}}
\delta_{\lam'_{s_3}\lam_{s_3}},
\nonumber\\
 && B^{(2)}_{\{\lam'_{s_k}\lam'_{t_k}\lam_{s_k}\lam_{t_k}\}} \equiv
\Bigl[
D^{(1/2)}_{\lam'_{s_1}\lam_{s_1}}(L^{-1}_\bK,\p_1)
D^{(1/2)}_{\lam'_{s_2}\lam_{s_2}}(L^{-1}_\bK,\p_2)
D^{(1/2)}_{\lam'_{s_3}\lam_{s_3}}(L^{-1}_\bK,\p'_3)
\Bigr]
\delta_{\lam'_{t_1}\lam_{t_1}}
\delta_{\lam'_{t_2}\lam_{t_2}}
\delta_{\lam'_{t_3}\lam_{t_3}},
\nonumber\\
 && B^{(3)}_{\{\lam'_{s_k}\lam'_{t_k}\lam_{s_k}\lam_{t_k}\}} \equiv
\Bigl[
D^{(1/2)}_{\lam'_{s_1}\lam_{s_1}}(L_{\p_{12}},\pmmo)
D^{(1/2)}_{\lam'_{s_2}\lam_{s_2}}(L_{\p_{12}},\pnno)
\Bigr]
\delta_{\lam'_{t_1}\lam_{t_1}}
\delta_{\lam'_{t_2}\lam_{t_2}}
\delta_{\lam'_{t_3}\lam_{t_3}}
\delta_{\lam'_{s_3}\lam_{s_3}},
\nonumber\\
&&  e j_{\mu}^{(3)\,\lam'_{s_3}\lam_{s_3}}
(\p'_3,\p_3)
\delta_{\lam'_{t_3}\lam_{t_3}}
 \equiv
 \la\p'_3\lam'_{s_3}\lam'_{t_3}|\hat{j}^{(3)}_{\mu}(0)|
 \p_3\lam_{s_3}\lam_{t_3}\ra.
\nonumber
\een
Here $j_{\mu}^{(3)}(0)$ is a current of third quark.
For convenience the electron charge $e$ is written as factor.
Sum of 3-momenta in the wave function of the relative motion
$\varphi$ is equal to zero:
$\po_1+\po_2+\po'_3 = 0$,
$\p_1+\p_2+\p_3 = 0$.
The list of notations is given at the end of the section.
The current normalization condition
$J_0(\0,\0)=e_N$
($e_N$ is the nucleon charge in units of the electron charge $e$)
leads to the WF normalization
condition~\cite{IzvVuzov1_Ilicheva}, which is given by:
\ben
\int
d\Omega_{\p_1}d\Omega_{\p_2}
\mid
\varphi_M(M_0)
\mid^2
/E_{\p_3}=1.
\label{norm_varphi}
\een
\par
The Dirac form factors of the nucleon are defined
by the expression:
\be
 J_{\mu}^{\lam'_J\lam_J}(\bK,\0) =
 {\dsty e_N} \bar U^{\lam'_J}(\bK)
 \{ \gamma_{\mu}F_1(Q^2)
 + \frac{i\kappa}{2M} F_2(Q^2)
 \sigma_{\mu\nu} q_{\nu}
         \}
 U^{\lam_J}({\0}),
 \label{ffactor def}
\ee
where $\kappa$ is the nucleon anomalous magnetic moment,
$
\sigma_{\mu\nu} = \frac{i}{2}(\gamma_{\mu}\gamma_{\nu} -
\gamma_{\nu}\gamma_{\mu})
$,
$Q^2\equiv -q^2>0$, $q=(P'-P)$, $P'=(E^M_{\bK}, \bK)$, $P=(M,\0).$
Sachs form factors is convenient for analysis
of the experimental data:
$G_E(Q^2) = F_1(Q^2) + F_2(Q^2)\kappa Q^2/4M^2$,
$G_M(Q^2)= F_1(Q^2) + \kappa F_2(Q^2)$.
Let write the zero component of the third quark current
like~(\ref{ffactor def}), but in form of Pauly $\sigma-$matrix,
with zero quark anomalous magnetic moments:
\ben
j_0^{(3)}(\p'_3,\p_3) =
 \frac{\hat{e}_q f_1(Q_{(3)}^2)}{2\sqrt{(E_3+m)(E'_3+m)}}
 [(E_{\p_3}+m)(E_{\p'_3}+m)
 + (\sigma\p'_3)(\sigma\p_3)],
\label{j_0}
\een
where $\hat{e}_3$ is charge operator of the third quark
in units of the electron charge,
$f_1$ is the quark Dirac form factor,
depended on $Q_{(3)}^2=-(p'_3-p_3)^2$.
Quark radius is defined by decomposition over $Q_{(3)}^2$:
$f_k(Q_{(3)}^2)\approx 1-\la r_{kq}^2\ra Q_{(3)}^2/6$.
\par
In work~\cite{IzvVuzov1_Ilicheva}
the relativistic three-particle oscillator was proposed
in the effective mass approximation.
This approximation is defined by the condition:
$m^{eff}_k- \la E_{\p_k}\ra<<m^{eff}_k$,
where $m^{eff}_k=\sqrt{m^2+\la p^2_k\ra}$,
$\la p^2_k\ra=\int
d\Omega_{\p_1}d\Omega_{\p_2}
|
\varphi_M
|^2
p^2_k
/E_{\p_3}$,
$\la E_{\p_k}\ra=\int
d\Omega_{\p_1}d\Omega_{\p_2}
|
\varphi_M
|^2
E_{\p_k}
/E_{\p_3}$.
The wave function $\varphi$ of relative motion
in this approximation is written as:
\be
\varphi_M^{osc}(\{\po_k\})
= N\exp[-\frac{m}{\gamma^2}(M_0-3m)]
\approx N\exp(-\frac{\bk^2}{\gamma^2}-\frac{3\bk'^2}{4\gamma^2})
\approx
\label{wf_osc_1}
\ee
$$\approx N \exp(-\frac{1}{\gamma^2}
(\po_1^2 + \po_2^2 + \po_1\po_2)\frac{2m}{m+m^{eff}}).$$
where
$ {\bf k} = ({\bf \pi}_1 - {\bf \pi}_2)/2 $,
$ {\bf k}' = (\pi_1 + \pi_2)/3 - 2\pi_3/3 $;
$\pi_k = \po_k\sqrt{2m/(m+E_{\po_k})}$ is
half-momentum~\cite{Skachkov_Solovcov78};
the energy has the form
$E_{\po_k} = m + \pi_k^2/2m$.
In the effective mass limit
  $\pi_k \approx \po_k\sqrt{2m/(m + m^{eff})}$,
$\pi_1 + \pi_2 + \pi_3 \approx 0$,
$E_{\po_k} \approx m + \po_k^2/(m + m^{eff})$.
If $\la p^2_k\ra$ is small, then $m^{eff} \approx m$
and effective mass approximation passes to
nonrelativistic limit.

Here we give a list of notations:
$d\Omega_{\p_k}=d\p_k/E_{\p_k}$,
$E_{\p_k}=\sqrt{\p_k^2 + m^2}$,
$E^M_K=\sqrt{\bK^2+M^2}$,
$\bK$ and $\p_k$ are nucleon and quark momenta,
$M$ and $m$ are nucleon and quark masses,
$U$ and $u$ are nucleon and quark spinors,
$\{\p_k \lam_{s_k}\lam_{t_k}\}\equiv
\p_1\lam_{s_1}\lam_{t_1}
\p_2\lam_{s_2}\lam_{t_2}
\p_3\lam_{s_3}\lam_{t_3};$
$\chi$ is SU(3) nucleon wave function.
The total spins and isospins of the nucleon and constituent quarks
are omitted and equal to $1/2$;
$\lam_J$, $\lam_{s_k}$ are the third projections
of the nucleon and quark spins and equal to $+1/2$.
We assume a sum over the repeated indices of
the third projection of spin
and isospin.
The normalization condition of vector states and
spinors of nucleon and quarks are given by:
$$
\la K' J' \lam'_J T' \lam'_T
\mid K J \lam_J T \lam_T\ra
=
(2\pi)^3
\delta^3(\bK'-\bK)
\delta_{\lam'_J \lam_J}
\delta_{\lam'_T \lam_T},
$$
$$
\la
\p'_k s_k \lam'_{s_k} t_k \lam'_{t_k}
\mid
\p_k s_k \lam_{s_k} t_k \lam_{t_k}
\ra
=
(2\pi)^3
E_{\p_k}
\delta^3(\p'_k-\p_k)
\delta_{\lam'_{s_k} \lam_{s_k}}
\delta_{\lam'_{t_k} \lam_{t_k}},
$$
$$
\bar{u}^{\lam_{s_k}}(\p)
u^{\lam_{s_k}}(\p)
=
m,
\;\;
\bar{U}^{\lam_J}(\bK)
U^{\lam_J}(\bK)
=
M/E^M_{\bK}
\mbox{ (there is no sum over } \lam_{s_k} \mbox{ and } \lam_J).
$$
$\stackrel{\not\,\,\not{o}}\p_k=L^{-1}_{\p_{12}}p_k
$, where $p_{12}=p_{1}+p_{2}$;
$\poc_k=(L^{-1}_K p_k)$,
$\ponotc_k=(L^{-1}_{\ponotc_{12}}\poc_k)$,
where $\poc_{12} = \poc_1+\poc_2$;
$L^{-1}_K$ is the Lorentz boost to the
rest frame of the nucleon and $L^{-1}_{\ponotc_{12}}$
and $L^{-1}_{p_{12}}$
are the Lorentz boosts to the
rest frame of [1+2] quarks.
Wigner rotation matrix has the form
(see~\cite{IzvVuzov1_Ilicheva} and references there in):
\be
D^{(1/2)}(L^{-1}_K, p_k)=
\frac
{(E_{\p_k}+m)(E^M_K+M)-(\sigma\bK)(\sigma\p_k)}
{\sqrt{
2(E_{\p_k}+m)(E^M_K+M)(E_{\bK} E_{\p_k}-\bK\p_k+mM)
}}.
\label{wigner_rotate}
\ee
4-momentum $\poc_k$ in the c.m.s. has the components:
\be
E_{\po_k}=\frac{1}{M}(E^M_{\bK}E_{\p_k}-\bK\p_k),\quad
\po_k=\p_k-\frac{\bK}{M}(E_{\p_k}-\frac{\bK\p_k}{M+E^M_{\bK}}).
\label{p_s_nulyom}
\ee

\begin{center}
{\bf 3. Analitycal analysis.}
\end{center}
\par
By using the current~(\ref{j_0}), the
expression for $G_E(Q^2)$ is represented as:
$G_E(Q^2)=\sqrt{2E^M_{\bK}/(M+E^M_{\bK})}J_0(\bK,\0).$
Electric nucleon radius, defined according to
the decomposition
$G_E(Q^2)\approx e_N -\la r^2_E\ra Q^2/6$, expresses as:
\be
\la r^2_E\ra=-{\bf\nabla}^2_{\bK}G_E(Q^2)|_{\bK=0}=
-\frac{3 e_N}{4M^2}-{\bf\nabla}_{\bK}^2 J_0(\bK,\0)|_{|{\bf K}=0}.
\label{r2_J}
\ee

  In the nonrelativistic limit we have from~(\ref{r2_J})
  and~(\ref{current}), denoting 
  $e_N=3(\chi,e_q^{(3)}\chi)$:
\be
\la r^2_E\ra^{NR} = e_N\int d\p_1 d\p_2
(-i{\bf \nabla}_{\bf K})^2\varphi(\po_1,\po_2,\po'_3)
\varphi(\p_1,\p_2,\p_3).
\label{r2_Nonrel_1}
\nonumber
\ee
Using the last expression of~(\ref{wf_osc_1})
with $m^{eff} \approx m$,
we have from~(\ref{r2_Nonrel_1}):
\ben
\la r^2_E\ra^{NR} = e_N\int d\p_1 d\p_2
|\varphi(\p_1,\p_2,\p_3)|^2\left[ -\frac{\p_3^2}{\gamma^4}
                                  + \frac{2}{\gamma^2} \right].
\label{r2_Nonrel_2}
\een
Now if we return to variables ${\bf k}$ and ${\bf k}'$,
assuming
$\p_1 = {\bf k} + {\bf k}'/2$, $\p_2 = -{\bf k} + {\bf k}'/2$
and using expression
for the relative motion wave function~(\ref{wf_osc_1}) 
in variables ${\bf k}$ and ${\bf k}'$,
we have well-known results~\cite{Petrunkin81}:
$
\la r^2_E \ra^{NR} = e_N/\gamma^2,
$
where ortonormalization condition~(\ref{norm_varphi}) is used also.
Now we can rewrite general expression for radius in the 
relativistic case with the oscillator 
wave function~(\ref{wf_osc_1}):
$$
\la r^2_E \ra = -\frac{3 e_N}{4M^2} +
\int d\Omega_{\p_1} d\Omega_{\p_2}
|\varphi|^2\left[
  -e_N\frac{m^2}{\gamma^4}({\nabla_{\bK}}\Mo'_0)^2
  \frac{1}{E_{\p_3}}
         + e_N\frac{m}{\gamma^2}
	 {\bf\nabla_{\bK}}^2\Mo'_0 \frac{1}{E_{\p_3}}
	  \right.+
$$
\be
\left.
	 + 3 \frac{m}{\gamma^2} ({\bf\nabla_{\bK}}\Mo'_0)
	 ({\bf\nabla_{\bK}} G_0)
	 -3 {\bf\nabla_{\bK}}^2 G_0
	 \right]_{|\bK=0}=
e_N [\frac{-3}{4M^2} + \la r^2_{I}\ra +
\la r^2_{II}\ra + \la r^2_{III}\ra + \la r^2_{IV}\ra].
\label{r2_Rel_1}
\ee
  Four terms $\la r^2_{I,II,III,IV}\ra$ correspond to four
expressions under integral in the expression~(\ref{r2_Rel_1}).
After differentiation over $\bK$ for $\bK=0$
we don't have any $\sigma$-matrix, that is why all expressions
are proportional
to the nucleon charge $e_N=3(\chi,e_q^{(3)}\chi)$.
So, in the quark model the  neutron radius equal to zero, but
neutron form factors deviate from zero
for unzero transfer momenta~\cite{Ilichova_Gomel2003}.
Further we consider only the proton radius.
Expressions for $\la r^2_{I}\ra$ and $\la r^2_{II}\ra$ have the forms:
\be
\la r^2_I \ra = -\frac{1}{M^2} \frac{m^4}{\gamma^4}
  \int d\Omega_{\p_1} d\Omega_{\p_2} |\varphi|^2\frac{1}{E_{p_3}}
   \frac{[\sum_{i=1}^3 E_{p_i}]^2}{E^2_{p_3}}\frac{\p_3^2}{m^2}
\label{r2_I}
\ee
\be
 \la r^2_{II} \ra = \frac{1}{M^2} \frac{m^2}{\gamma^2}
  \int d\Omega_{\p_1} d\Omega_{\p_2} |\varphi|^2\frac{1}{E_{p_3}}
\left[ 3(E_{p_1}+E_{p_2})
                +\frac{\p_{3}^2}{E^2_{p_3}}
                   (E_{p_3} - E_{p_1}-E_{p_2})
\right]
\frac{\sum E_{p_i}}{E_{p_3} m}
\label{r2_II}
\ee
Third term $\la r^2_{III}\ra$ contains derivative from
${G_0}$~(\ref{G_mu}). In this expression
$B^{(3)}$ does not depend on $\bK$.
It was shown numerically that expression
$\nabla_{\bK} B^{(1)}|_{\bK=0}$ is negledibly small
in the wide regionof the parameters
and
$\nabla_{\bK} B^{(1)}|_{\bK=0}$ is proportional to
$\pmmo + \pnno = 0$ in the effective mass limit.
Taking into account that $\p_1 + \p_2 = -\p_3$,
it is possible to show that contribution of $B^{(2)}$ is equal
to zero.
So, we have only contribution from $1/E_{\po'_3}$ and from the
quark current:
\ben
&&\la r^2_{III} \ra \approx \frac{1}{M^2} \frac{m^2}{\gamma^2}
  \int d\Omega_{\p_1} d\Omega_{\p_2} |\varphi|^2\frac{1}{E_{p_3}}
\Big[ E_{p_3} - E_{p_1} - E_{p_2} \Big]
  \frac{\sum E_{p_i}}{2 m E_{p_3}}\frac{\p_3^2}{E^2_{p_3}}
\label{r2_III_1}
\een
Nonrelativistic expression for radius $(e_N/\gamma^2)$ 
can be obtained
from~(\ref{r2_I})-(\ref{r2_III_1}),
assuming $M \approx 3m$.
\par
Relativistic term $\la r^2_{IV} \ra$ is represented as
series of the negative powers  $[m/\gamma]^2$
(including the zero power).
As a result we have:
\ben
\la r^2_E \ra  &=& \frac{9}{M^2} \Big(
  \frac{m^4}{\gamma^4} a_{-4} +\frac{m^2}{\gamma^2} a_{-2}
  + a_0 + \frac{\gamma^2}{m^2} a_{2}  + 
  \frac{\gamma^4}{m^4} a_{4}  + ...
                              \Big).
\label{r2_serie}
\een
Therefore we don't have dimensional terms in the form $1/m^2$
for radius~(\ref{r2_serie})
 (compare~\cite{Petrunkin81, Fajimara_70}).

 Dependence on the nucleon momentum $\bK$ in~(\ref{current})
 is defined by Lorentz-transformation, which contains 3-velosity
$\bK/M$ (see~(\ref{p_s_nulyom}) ).
Therefore, the nucleon radius~(\ref{r2_J})
for arbitrary wave function is defined as $f(m/\gamma)/M^2$,
where $f$ is undimensional function.If quarks have structure, 
then $f$ depends on the parameters of the
quark structure.
\begin{center}
{\bf 3. Numerical analysis.}
\end{center}
\par
Figure $a$ shows dependences of proton radius
$\la r^2_E \ra$ (curve 1),
$\la r^2_{I} \ra +\la r^2_{II} \ra +\la r^2_{III} \ra$ (curve 2)
and $\la r^2_{IV} \ra$ (curve 3)
from $m/\gamma$.
Points for $\la r^2_E \ra$ was obtained as follows:
data for current and for electric form factor $G_E(Q^2)$
were obtained for fixed $m/\gamma$ and in the interval
$0.001<Q^2<0.1$ GeV$^2$.
Integral was calculated by VEGAS~\cite{Lepage78} with five iteration
and 10$^6$ trials in each.
After that points of $G_E(Q^2)$ were fitted by function
$G_E(Q^2) = b_0 - b_1Q^2/6  + b_2 Q^4 +  b_3 Q^6 +  b_4 Q^8 $
for obtaining the radius $\la r^2_E \ra \equiv b_1$.
Errors of the fit are shown on the figures.

For calculation of
$\la r^2_{I} \ra +\la r^2_{II} \ra +\la r^2_{III} \ra$
we can use~(\ref{r2_I})-(\ref{r2_III_1}), calculated
by VEGAS~\cite{Lepage78}. Error of the calculation is small and shown
on the figure.
For calculation of $\la r^2_{IV} \ra$ it was difficultly
to obtain analitycal formulae, 
like~(\ref{r2_I})-(\ref{r2_III_1}).
However,
we can separate the contribution of
$\la r^2_{IV} \ra$ in the current~(\ref{current}).
For this it is sufficiently to replace
$\varphi(\Mo'_0)\to\varphi(M_0)$ and perform calculation as in
case of $\la r^2_E \ra$.
\par
If $m/\gamma \to \infty$ then $\la r^2_E \ra$ increases
because of there are positive powers of $m/\gamma$,
as in nonrelativistic theory.
If $m/\gamma$ decreases then relativistic term with
negative powers of $m/\gamma$ dominates.

From figure $a$ we can see that
$\la r^2_{I} \ra +\la r^2_{II} \ra +\la r^2_{III} \ra$
has nonzero limit for $m/\gamma \to 0$.
Indeed, it is possible to represent integrals
in~(\ref{r2_I}),~(\ref{r2_II}),~(\ref{r2_III_1})
as infinite series of the decomposition with the negative
powers of  $m/\gamma$. This decompositions have nonzero
terms $[m/\gamma]^{-4}$ in~(\ref{r2_I}) and $[m/\gamma]^{-2}$
in~(\ref{r2_II}) and~(\ref{r2_III_1}), which define limits
$m/\gamma \to 0$ for
$\la r^2_I\ra$, $\la r^2_{II}\ra$ É $\la r^2_{III}\ra$.
Nonzero limit of strong coupling for the proton radius
is relativistic effect. In the nonrelativistic case
size of system decreases to zero when the bound energy increases.

From figure $a$ we can see, that behaviour of the radius
of the bound system is more complicated in the limit of strong coupling
as compare to the behaviour of
$\la r^2_{I} \ra +\la r^2_{II} \ra +\la r^2_{III} \ra$.
In the region of small $m/\gamma$ form factor $G_E(Q^2)>1$
for small $Q^2$, radius of the bound system is negative and its
module increases when $m/\gamma \to 0$.
We can analyze what terms in the expression~(\ref{current})
correspond to this behaviour of the form factor in the
ultrarelativistic region. From figure $a$ we can see that the
cause is the term $\la r^2_{IV}\ra$.
According to~(\ref{current}),~(\ref{G_mu})
contribution of $\la r^2_{IV}\ra$ consist of following parts:
contribution of the Wigner rotation matrix $\la r^2_{D}\ra$ and
contribution of the quark current $\la r^2_{j}\ra$.
Cross term in which first derivative ${\bf\nabla}_{\bK}$ acts to
Wigner rotation matrix, second derivative acts
to the quark current, does not give any contribution
(see explanation before~(\ref{r2_III_1})).
Figure $b$ shows the proton radius dependences on
$m/\gamma$:
$\la r^2_E \ra$ (curve 1), $\la r^2_{j}\ra$ (curve 4),
$\la r^2_{D}\ra$ (curve 5).
It is possible to conclude, that the positive peak and the
negative proton radius for small $m/\gamma$ 
is originated from the quark current.
So, the reason for the negative radius of the bound system of the
spinor particle is the dependence of the valence quark current
on quark momenta, defined by expression~(\ref{j_0}).

\begin{figure}
\epsfig{file=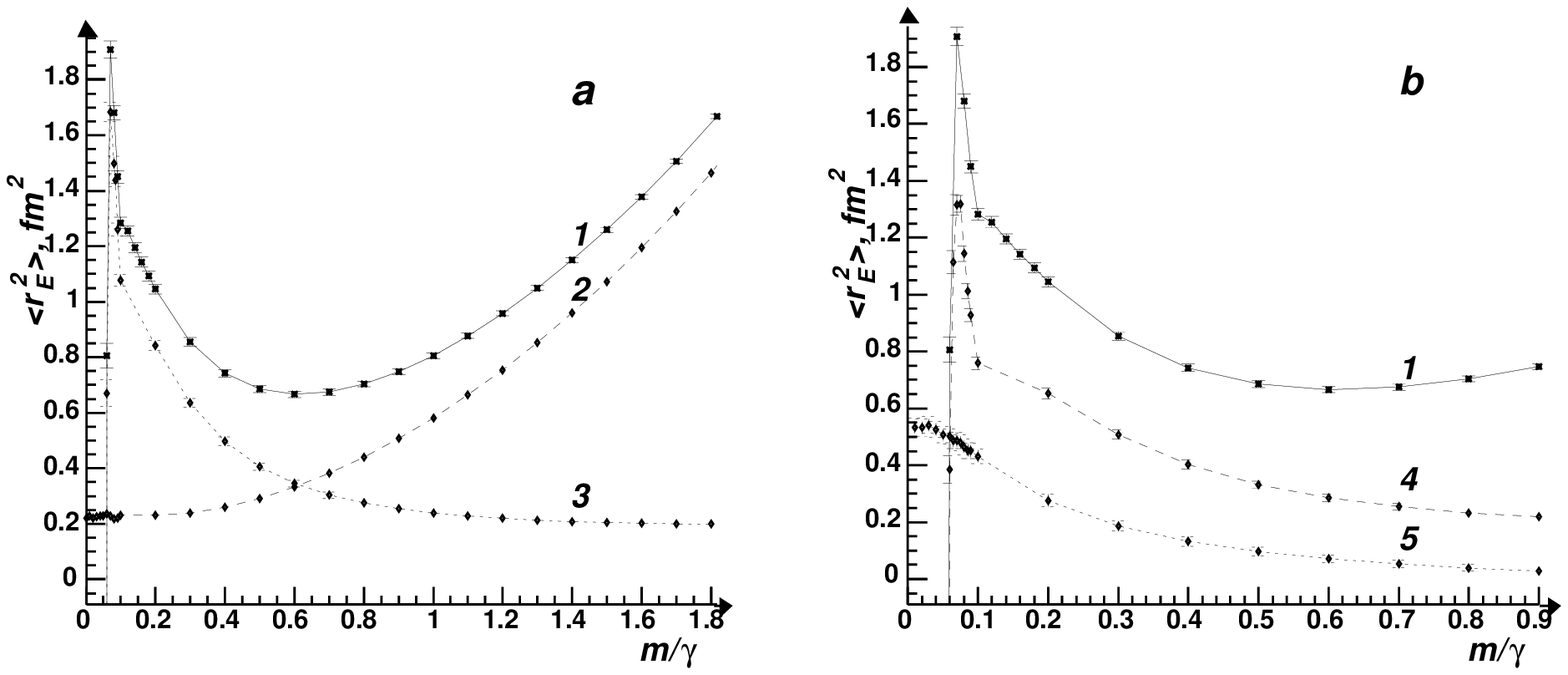,width=16cm}
\label{fig}
\end{figure}

The quark radius contributes to the terms with negative powers
of $m/\gamma$ in the expression~(\ref{r2_serie}).
We can write down the quark radius contribution to the proton
radius exactly.
\ben
 \Delta\la r^2_E\ra =
 \frac{9}{M^2}
 \la r^2_{1q} \ra m^2
 \int d\Omega_{\p_1} d\Omega_{\p_2} |\varphi|^2\frac{1}{E_{p_3}}
    \frac{m^2 + 2 E_{p_3}^2}{3 E_{p_3}^2}
    \Big(\frac{\sum_{i=1}^{3} E_{p_i}}{3m}\Big)^2
\label{Delta_r2_E}
\een
\par
In work~\cite{IzvVuzov2_Ilicheva} for oscillator
model~(\ref{wf_osc_1}) it was shown that simultaneous
description of proton and
neutron magnetic moments with 2$\%$ accuracy is obtained for
$m/\gamma=1.818$.
For this value proton radius is more than experimental data
$\sqrt{\la r^2_E\ra}=0.870\pm0.008$ fm~\cite{Particle_Data_2004}
and is equal to $\sqrt{\la r^2_E\ra}=1.291\pm0.003$ fm.
For simultaneous description of experimantal data
it is need to include negative contribution into the
proton mean square radius, which is equal to
$\Delta\la r^2_E\ra  = -0.910$ fm$^2$
for experimental data $\sqrt{\la r^2_E\ra}=0.870\pm0.008$ fm.
Assuming that this contribution is related to
the valence quark radius, we have
$\la r^2_{1q} \ra m^2 = -1.875$.

Notice, that nonzero quark radius does not influence 
on the nucleon magnetic moments, by using of which the ratio
$m/\gamma=1.818$ was obtained.
If quark does not have size, then
value, obtained in this model is corresponding to experimental data
for
$m/\gamma = 0.4$ and $m/\gamma = 0.9$ (see figure a).
For this $m/\gamma$
it is possible to
describe the nucleon magnetic moments but with less accuracy
($m/\gamma = 0.4$: deviations are 11$\%$ for proton and
 20$\%$ for neutron),
($m/\gamma = 0.9$: deviations are 4.5$\%$ for proton and
11$\%$ for neutron);
or in other case
if we introduce large the quark anomalous magnetic moments,
which should be different for $u$- and 
$d$-quarks~\cite{IzvVuzov2_Ilicheva}.
So, if we assume that good description of the proton
and neutron
magnetic moments with the zero quark anomalous magnetic moments
is not accidental, then model indicates that the valence
quarks must have the negative radius.
\begin{center}
{\bf Conclusion.}
\end{center}
\par
So, in this model the nucleon radius is defined by
undimensional parameters
$m/\gamma$ and $\la r^2_1q \ra m^2$.
Absolute values of the quark characteristics (the quark mass
and the quark radius) are not defined without specification 
of the "scale"-parameter, which defined the coupling energy.
Oscillator wave function was used here for calculation,
but the radius
expression~(\ref{r2_serie}) with factor $1/M^2$
is general; it is possible that
additional powers of $m/\gamma$ can change
series~(\ref{r2_serie})
for other form of wave function and
other definition of parameter $\gamma$.
Dimensional factor $1/M^2$ is consequence of relativistic
kinematic of the relative quark motion
(see~(\ref{p_s_nulyom}) and~(\ref{r2_J})).
Scale invariant dependence of the nucleon radius and magnetic
moments on $m/\gamma$ is related to that the wave function and quasipotential
are independent on bound system mass $M$ in the impulse
aproximation~\cite{IzvVuzov1_Ilicheva}.
In the general case the relativistic expression for radius
of the bound system contains infinite series of negative
powers of $m/\gamma$. In the oscillator model this
gives large negative radius of the system with
strong coupling of the
spinor particles.
Wigner rotation matrices gives small contributions into
the proton radius in the region of parameters, which
correspond to the the experimental data of the proton radius.
In this model it is need to intoduce
negative quark radius for simultaneous
description of the proton radius and nucleon magnetic moments.

\end{document}